# Assembled Kinetic Impactor for Deflecting Asteroids via Combining the Spacecraft with the Launch Vehicle Final Stage

Yirui Wang[1,2], Mingtao Li*[1,2], Zizheng Gong[3], Jianming Wang[4], Chuankui Wang[4], Binghong Zhou[1,2]

[1] National Space Science Center, Chinese Academy of Sciences

[2] University of Chinese Academy of Sciences

[3] Beijing Institute of Spacecraft Environment Engineering, China Academy of Space Technology

[4] Beijing Institute of Astronautical Systems Engineering, China Academy of Launch Vehicle Technology

*Correspondence Author E-mail: limingtao@nssc.ac.cn

## Abstract

Asteroid Impacts pose a major threat to all life on the Earth. Deflecting the asteroid from the impact trajectory is an important way to mitigate the threat. A kinetic impactor remains to be the most feasible method to deflect the asteroid. However, due to the constraint of the launch capability, an impactor with the limited mass can only produce a very limited amount of velocity increment for the asteroid. In order to improve the deflection efficiency of the kinetic impactor strategy, this paper proposed a new concept called the Assembled Kinetic Impactor (AKI), which is combining the spacecraft with the launch vehicle final stage. That is, after the launch vehicle final stage sending the spacecraft into the nominal orbit, the spacecraft-rocket separation will not be performed and the spacecraft controls the AKI to impact the asteroid. By making full use of the mass of the launch vehicle final stage, the mass of the impactor will be increased, which will cause the improvement of the deflection efficiency. According to the technical data of Long March 5 (CZ-5) launch vehicle, the missions of deflecting Bennu are designed to demonstrate the power of the AKI concept. Simulation results show that, compared with the Classic Kinetic Impactor (CKI, performs spacecraft-rocket separation), the addition of the mass of the launch vehicle final stage can increase the deflection distance to more than 3 times, and reduce the launch lead-time by at least 15 years. With the requirement of the same deflection distance, the addition of the mass of the launch vehicle final stage can reduce the number of launches to 1/3 of that of the number of CKI launches. The AKI concept makes it possible to defend Bennu-like large asteroids by a no-nuclear technique within 10-year launch lead-time. At the same time, for a single CZ-5, the deflection distance of a 140 m diameter asteroid within 10-year launch lead-time, can be increased from less than 1 Earth radii to more than 1 Earth radii, which means the improvement of the reliability and efficiency of asteroid deflection missions.

# Introduction

Scientific interest in Near-Earth Objects (NEOs) is great because it seems that many of these objects are main-belt asteroids which have been perturbed into terrestrial-planet-crossing orbits, thus constituting a large proportion of the impactors on terrestrial-planet surfaces[1,2]. Several serious impact events (i.e. Chixulub event[3], Tunguska event[4] and Chelyabinsk event[5]) have aroused people's attention to the threat of the asteroid impacts. Planetary defense is to use active methods (nuclear explosion, kinetic impactor, laser ablation, gravitational tractor, etc.) to destroy the structure of the asteroid or to deflect the orbit of the asteroid. Despite of fragmentation risks, a kinetic impactor remains a promising strategy for asteroid deflection[6]. In 2005, the Deep Impact mission released an impactor weighing 372 kg to collide with comet Tempel 1 at a velocity of 10.2 km/s[7]. This impact generated a 0.0001 mm/s velocity increment in the comets orbital velocity and decreased its perihelion distance by 10 meters[8]. NASA and ESA have cooperated in the AIDA (Asteroid Impact and Deflection Assessment) mission, of which DART (Double Asteroid Redirection Test) will be launched in July 2021, with a 555 kg spacecraft hitting Didymos' moon at a relative velocity of 6.65 km/s, generating a velocity increment of 0.8-2 mm/s (depending on the ejecta enhancement factor $\beta$, which measures an additional momentum transferred to the target by the thrust in the opposing direction of crater ejecta that escapes the gravitational attraction of the target body[9]). Although the kinetic impactor strategy is currently the most feasible method to deflect asteroids, the constraints of the launch capability have resulted in a limited deflection efficiency. Barbee, et al. [10] design the Hypervelocity Asteroid Mitigation Mission for Emergency Response (HAMMER) for deflecting asteroid Bennu. The simulation results show that in order to deflect Bennu for 1.4 Earth radii, at least 17 Delta IV Heavy launch vehicles are needed with a 25-year launch lead-time, 75 Delta IV Heavy launch vehicles are needed with a 10-year launch lead-time. Such a large number of launches will significantly increase the risk of the failure of the asteroid deflection missions. Many studies have been performed to improve the deflection efficiency of the kinetic impact strategy: H-reversal orbit[11] can greatly enhance the impact velocity to 90 km/s, but it's technically immature to deliver a solar sail into such high-energy retrograde orbits at the current stage. Enhanced Kinetic Impactor (EKI)[12] is proposed to increase the impactor mass by maneuvering space rocks, which can increase the deflection distance by one order of magnitude compared to Classic Kinetic Impactor (CKI). However, some key technologies for EKI, such as the asteroid capture, need to be further verified in orbit.

For a general space mission, the spacecraft-rocket separation is performed after the launch vehicle final stage is sending the spacecraft into the nominal orbit. The reason for the separation is that the launch vehicle final stage will become useless after the spacecraft enters the nominal orbit. However, for special space missions such as kinetic impact missions, the launch vehicle final stage will not become useless. That's to say, the launch vehicle final stage can be used as a payload to improve the mass of the impactor, thereby improving the deflection efficiency of the kinetic impactor strategy.

This paper proposes a new concept of the kinetic impact strategy, which is called the Assembled Kinetic Impactor (AKI) that combined the spacecraft with the launch vehicle final stage. That is, after the launch vehicle final stage sending the spacecraft into the Earth escape orbit, the spacecraft-rocket separation will not be performed and the spacecraft controls the AKI to impact the asteroid. By making full use of the mass of the launch vehicle final stage, the impactor's mass will be increased, which will cause the improvement of the deflection efficiency. Taking deflecting Bennu as an example, and referring to the technical parameters of the Long March 5 launch vehicle, the AKI missions are designed. At the same time, the technical feasibility of the AKI is preliminarily discussed.

# Results

The Long March 5 (CZ-5) is currently the largest and most powerful launch vehicle in China. Its launch performance is similar to Delta IV heavy, Proton K and Ariane 5. CZ-5's TMI (Trans-Mars injection) orbit launch capability is 5~6 t. The conventional CZ-5 adopts a two-and-a-half configuration, the final stage mass is 7.3 t, of which the dry mass is about 6.5 t and the fuel redundancy is about 800 kg. In general space missions, the redundant fuel should be discharged for avoiding explosions. Therefore, the redundant fuel can't be used to increase the mass of the impactor. In order to make full use of the mass of the redundant fuel. The launch vehicle final stage in the AKI doesn't carry the redundant fuel, and the normally equipped 800 kg redundant fuel is carried by the spacecraft.

101955 Bennu (1999 RQ36) is considered to be one of the Potential Hazardous Asteroids (PHA), whose diameter is 492 m with the mass of $7.9 \times 10^{10}$ kg[13]. It is target of the OSIRIS-Rex sample return mission. According to the orbital prediction results of JPL Horizons On-Line Ephemeris System[14], Bennu makes a very close approach to the Earth on September 25, 2135, with the minimum geocentric distance of 0.00199AU[15]. This paper refers to HAMMER mission design method[10], taking Bennu's minimum geocentric distance (which is on September 25, 2135) as the research object, the following 2 objective functions are used to demonstrate the power of the AKI missions: 1) The maximum deflection distance of a single AKI with a 10-year launch lead-time; 2) The minimum number of AKI launches for the requirement to achieve a deflection of at least 1.4 Earth radii with a 10-year launch lead-time.

**The maximum deflection distance of a single AKI**

With a 10-year launch lead-time, assuming $\beta=1$, for the objective function of the maximum deflection distance of a single AKI, the simulation results are shown in Table 1, which includes the corresponding results of a single Classic Kinetic Impactor (CKI, performs spacecraft-rocket separation). Figure 1 shows the AKI mission's transfer trajectory. According to the simulation results, the maximum deflection distance of a single CKI is 113.57 km, and the maximum deflection distance of a single AKI is 399.34 km. Compared with the CKI, the addition of the launch vehicle final stage mass can increase the deflection distance to more than 3 times. Meanwhile, it can be seen from the optimization results, the AKI trends to increase the impact velocity while increasing the impactor mass. This is because the spacecraft in AKI has a smaller mass (2.25 t), this smaller mass corresponds to higher launch $C_3$ (42.89 km²/s²), thereby achieving a higher impact velocity (7.17 km/s).

Table 1 10-year launch lead-time optimal deflection of a single impactor

| Type | AKI (With Final Stage) | CKI (Without Final Stage) |
| --- | --- | --- |
| Launch Vehicle | CZ-5 | CZ-5 |
| Number of Launches | 1 | 1 |
| $C_3$ | 42.89 km²/s² | 13.75 km²/s² |
| Impactor Mass | 8.75 t | 5.09 t |
| Spacecraft Mass | 2.25 t | 5.09 t |
| Launch Date | 2125-1-27 6:44:25 | 2125-1-13 9:7:34 |
| Flight Time | 1057.31 days | 651.14 days |
| Impact Date | 2127-12-20 14:9:41 | 2126-10-26 12:24:19 |
| Impact Velocity | 7.17 km/s | 4.15 km/s |
| Bennu $\Delta v$ | 0.79 mm/s | 0.27 mm/s |
| Deflection Time | 2836.82 days | 3256.89 days |
| Deflection Distance | 399.34 km | 113.57 km |

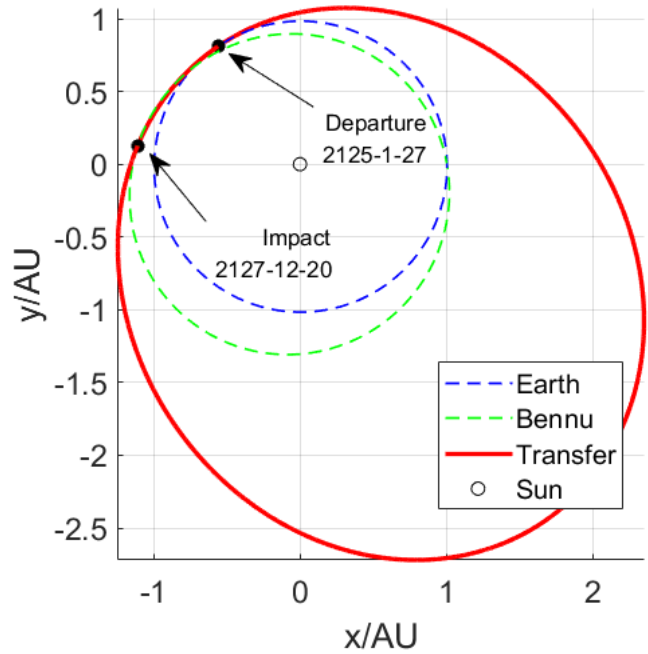

Figure 1 10-year launch lead-time optimal deflection trajectory

## The minimum number of AKI launches for the requirement to achieve a deflection of at least 1.4 Earth radii

With a 10-year launch lead-time, assuming $\beta=1$, for the objective function of the minimum number of AKI launches for the requirement to achieve a deflection of at least 1.4 Earth radii, the simulation results are shown in Table 2, which includes the corresponding results of the CKI. According to the simulation results, compared with the CKI, the addition of the launch vehicle final stage mass can reduce the required number of launches from 79 to 23 for the CZ-5, which can greatly reduce the cost of mission launches and improves the mission reliability. Introducing the simulation results of Barbee, et al. [10], under the same condition, the required number of launches is 75 for the Delta IV. Since the launch performance of the Delta IV and the Long March 5 is similar, the correctness of the algorithm in this paper can be verified.

Table 2 Deflection of at least 1.4 RE with 10-year launch lead-time and $\beta=1$

| Type | AKI (With Final Stage) | CKI (Without Final Stage) |
|---|---|---|
| Launch Vehicle | CZ-5 | CZ-5 |
| Number of Launches | 23 | 79 |
| $C_3$ | 43.00 km²/s² | 13.78 km²/s² |
| Impactor Mass | 200.96 t | 401.41 t |
| Launch Date | 2125-1-26 14:27:40 | 2125-1-12 1:6:5 |
| Flight Time | 1056.72 days | 651.65 days |
| Impact Date | 2127-12-19 7:45:45 | 2126-10-25 16:44:14 |
| Impact Velocity | 7.15 km/s | 4.15 km/s |
| Bennu $\Delta v$ | 18.18 mm/s | 21.08 mm/s |
| Deflection Time | 2838.08 days | 3257.71 days |

| Deflection Distance | 1.45 Re (9224.73 km) | 1.41 Re (8988.86 km) |

# Discussion

**Deflection efficiency of the Assembled Kinetic Impactor**

This section further discusses the influences on the deflection efficiency when the launch lead-time, $\beta$ and the asteroid size have changed. First, the influence of the launch lead-time on the deflection efficiency is discussed. With a 25-year launch lead-time, assuming $\beta=1$, for the objective function of the maximum deflection distance of a single AKI, the simulation results are shown in Table 3.

Table 3 25-year launch lead-time optimal deflection of a single impactor

| Type | AKI (With Final Stage) | CKI (Without Final Stage) |
| --- | --- | --- |
| Launch Vehicle | CZ-5 | CZ-5 |
| Number of Launches | 1 | 1 |
| $C_3$ | 45.27 km$^2$/s$^2$ | 20.03 km$^2$/s$^2$ |
| Impactor Mass | 8.55 t | 4.09 t |
| Spacecraft Mass | 2.05 t | 4.09 t |
| Launch Date | 2110-10-10 21:42:38 | 2111-10-14 21:45:40 |
| Flight Time | 980.06 days | 612.34 days |
| Impact Date | 2113-6-16 23:15:32 | 2113-6-18 5:59:15 |
| Impact Velocity | 5.80 km/s | 4.27 km/s |
| Bennu $\Delta v$ | 0.63 mm/s | 0.22 mm/s |
| Deflection Time | 8136.44 days | 8135.16 days |
| Deflection Distance | 874.10 km | 224.66 km |

Compared with the results of Table 1, the deflection distance (399.34 km) achieved by a single AKI with a 10-year launch lead-time is larger than the deflection distance (224.66 km) achieved by a single CKI with a 25-year launch lead-time. This means that the addition of the launch vehicle final stage mass can help to reduce the launch lead-time by at least 15 years.

Second, the influence of $\beta$ on the deflection efficiency is discussed. With a 10-year launch lead-time, assuming $\beta=2.5$, for the objective function of the minimum number of AKI launches for the requirement to achieve a deflection of at least 1.4 Earth radii, the simulation results are shown in Figure 4.

Table 4 Deflection of at least 1.4 RE with 10-year launch lead-time and $\beta=2.5$

| Type | AKI (With Final Stage) | CKI (Without Final Stage) |
| --- | --- | --- |
| Launch Vehicle | CZ-5 | CZ-5 |
| Number of Launches | 9 | 32 |
| $C_3$ | 42.96 km$^2$/s$^2$ | 13.76 km$^2$/s$^2$ |
| Impactor Mass | 78.66 t | 162.70 t |
| Launch Date | 2125-1-26 20:0:35 | 2125-1-13 2:19:49 |
| Flight Time | 1056.86 days | 651.54 days |
| Impact Date | 2127-12-19 16:38:28 | 2126-10-26 15:19:37 |
| Impact Velocity | 7.15 km/s | 4.15 km/s |
| Bennu $\Delta v$ | 17.81 mm/s | 21.36 mm/s |
| Deflection Time | 2837.71 days | 3256.77 days |
| Deflection Distance | 1.41 Re (9014.75 km) | 1.43 Re (9107.69 km) |

Compared with the results of Table 2, for the same requirement to achieve a deflection of at least 1.4 Earth radii, when the $\beta$ is 2.5, the required number of CKI launches is 32 for the CZ-5 (Introducing the simulation results of Barbee, et al. [10], under the same condition, the require number of launches is 30 for the Delta IV). If the AKI concept is adopted, the addition of the launch vehicle final stage mass can reduce the required number of launches from 32 to 9 for the CZ-5, which makes it possible to defend Bennu-like large asteroids by a no-nuclear technique within 10 years. With the development of the future heavy launch vehicle, the deflection efficiency of the AKI concept will be further improved.

Finally, the influence of the asteroid size on the deflection efficiency is discussed. Potentially Hazardous Asteroids (PHAs), which are the asteroids with a minimum orbit intersection distance (MOID) with Earth of less than 0.05 AU and a diameter of over 140 m, will cause significant regional damage in the event of impact. Such asteroids are the focus of the asteroid defense missions. Assuming that there is a virtual asteroid of 140 m diameter moving on the same orbit of Bennu (C-type and S-type are discussed for this virtual asteroid: C-type sphere has a density of 1.26 g/cm$^3$ and a mass of 1.81 x10$^9$ kg, S-type sphere has a density of 2.80 g/cm$^3$ and a mass of 4.02 x10$^9$ kg ), the deflection efficiency for this 140 m diameter PHA by a single AKI is calculated. With a 10-year launch lead-time, assuming $\beta$=1, for the objective function of the maximum deflection distance of a single AKI, the simulation results are shown in Table 5.

Table 5 Deflection of 140-m diameter asteroid with 10-year launch lead-time

| Asteroid Density | 1.26 g/cm$^3$ | | 2.80 g/cm$^3$ | |
| --- | --- | --- | --- | --- |
| Type | AKI | CKI | AKI | CKI |
| Launch Vehicle | CZ-5 | CZ-5 | CZ-5 | CZ-5 |
| Number of Launches | 1 | 1 | 1 | 1 |
| C$_3$ | 42.94 km$^2$/s$^2$ | 13.76 km$^2$/s$^2$ | 42.96 km$^2$/s$^2$ | 13.76 km$^2$/s$^2$ |
| Impactor Mass | 8.74 t | 5.08 t | 8.74 t | 5.08 t |
| Spacecraft Mass | 2.24 t | 5.08 t | 2.24 t | 5.08 t |
| Launch Date | 2125-1-26 18:33:17 | 2125-1-13 23:55:12 | 2125-1-26 20:16:46 | 2125-1-13 7:53:12 |
| Flight Time | 1057.26 days | 651.37 days | 1056.89 days | 651.51 days |
| Impact Date | 2127-12-20 0:43:30 | 2126-10-27 8:49:55 | 2127-12-19 17:37:1 | 2126-10-26 20:13:15 |
| Impact Velocity | 7.16 km/s | 4.15 km/s | 7.15 km/s | 4.15 km/s |
| Bennu $\Delta v$ | 34.57 mm/s | 11.65 mm/s | 15.54 mm/s | 5.24 mm/s |
| Deflection Time | 2837.38 days | 3256.04 days | 2837.67 days | 3256.56 days |
| Deflection Distance | 2.75 Re (17538.81 km) | 0.78 Re (4965.44 km) | 1.23 Re (7865.25 km) | 0.35 Re (2231.24 km) |

According to the simulation results, for this 140 m diameter virtual asteroid, using CZ-5 to launch a kinetic impactor with a 10-year launch lead-time. Regardless whether this asteroid is C-type or S-type, a single CKI can't achieve a deflection distance of 1 Earth radii, which cannot completely eliminate the threat of the asteroid impact. If a single AKI is used, the deflection distance of the C-type asteroid can reach 2.75 Earth radii, and the deflection distance of the S-type asteroid can reach 1.23 Earth radii. Compared with the CKI concept, the AKI concept can effectively improve the defense capability of the 140-m asteroids.

**Technical feasibility of the Assembled Kinetic Impactor**

In order to demonstrate the technical feasibility of the AKI, this section uses the design results of Table 1 (the total mass of the AKI is 8.75 t, of which the spacecraft mass is 2.25 t and the launch vehicle final stage is 6.5 t) as an example, the AKI is designed. An AKI is composed of a spacecraft and a launch vehicle final stage, and the spacecraft is composed of a propulsion module and an instrument module. The schematic diagram of the AKI is shown in Figure 2. The launch vehicle final stage is installed on the -x side of the spacecraft, and the propulsion module is installed on the +x side of the spacecraft.

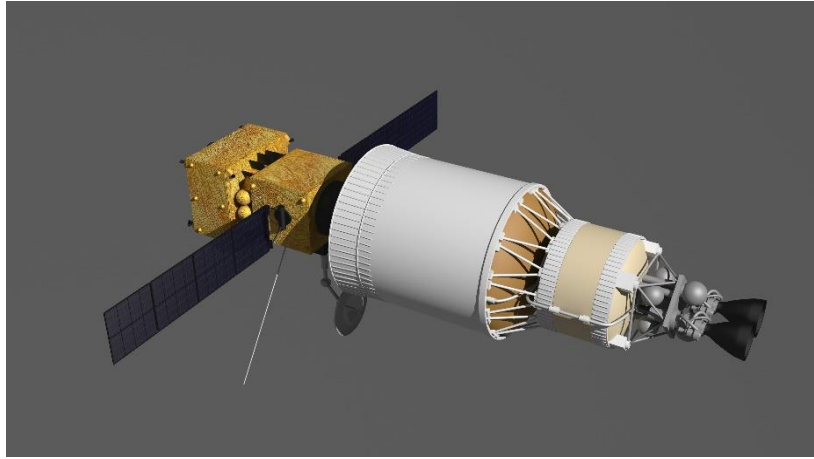

Figure 2 The schematic diagram of the Assembled Kinetic Impactor

The Long March 5 fairing has a diameter of 5.2 m and a height of 12.7 m. Figure 3 shows a schematic diagram of the AKI within the CZ-5 fairing. Taking into account the technical feasibility and the cost, the AKI concept doesn't consider the reactivation of the launch vehicle final stage to provide additional orbital maneuvers. In order to eliminate the explosion risks of the launch vehicle final stage during the mission, no redundant fuel is carried by the launch vehicle final stage. The launch vehicle final stage adopts a strategy of fuel-exhaustion-shutoff. If the launch $C_3$ provided by the launch vehicle final stage is insufficient, the monopropellant thrusters on the spacecraft can be used to correct the launch injection error and provide the velocity increment required for mid-correction. If the thruster specific impulse is 220 s, Figure 4 shows the relationship between the velocity increment and the fuel cost.

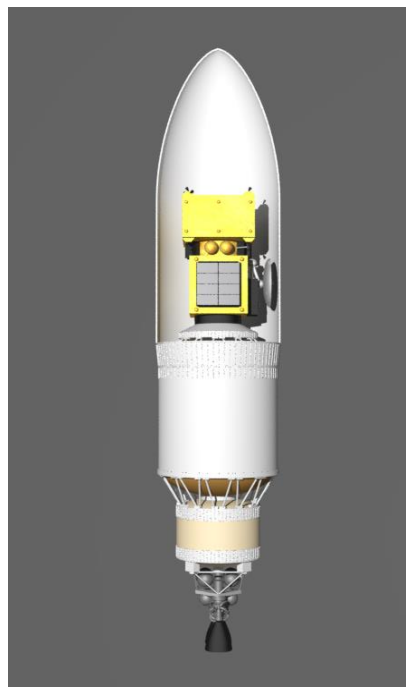

Figure 3 The Assembled Kinetic Impactor within the Long March 5 fairing

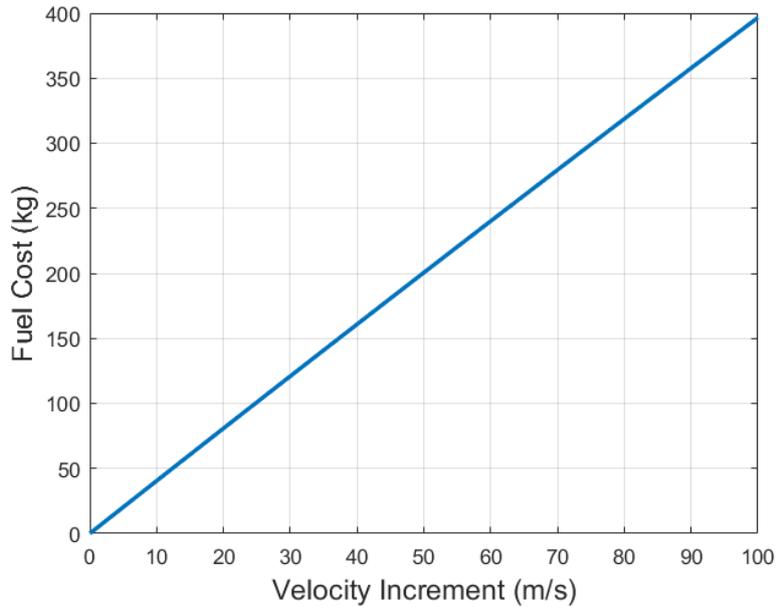

Figure 4 Fuel cost corresponding to the velocity increment of Assembled Kinetic Impactor

Compared with the CKI, the AKI concept poses new challenges to the attitude & orbit control subsystem, so this paper mainly introduces the design results of the attitude & orbit control subsystem. The main tasks of attitude & orbit control subsystem are as follows: 1) during the early phase of in-orbit, eliminate the attitude angle deviation, complete the attitude capture and establish the attitude toward the sun; 2) during the deep space transfer stage, complete the mid-correction; 3) during the approach process, complete the attitude control of the AKI, achieve the 3-axis stable pointing control, and meet the pointing requirements of the navigation sensor; 4) during the terminal guidance stage, complete the attitude & orbit control required for the terminal guidance of the accurate impact.

After the spacecraft and the launch vehicle final stage are assembled together, the main difficulties for attitude control are as follows: 1) in order to avoid the coupling of attitude control and orbit control, it is necessary to form force couples to control the attitude, which means the attitude control will not affect the orbit control. However, the center of mass of the AKI is located on the launch vehicle final stage, how to design the attitude control subsystem without modifying the launch vehicle final stage? At the same time, the solar arrays and launch vehicle final stage is behind the AKI's forward direction, how to prevent the thruster plumes from affecting the solar arrays and launch vehicle final stage? 2) the AKI is a slender body with a large moment of inertia, the moment of inertia of the launch vehicle final stage (approximately cylindrical: $I_x$=87880 kg·m², $I_y$= $I_z$ =156260 kg·m²) is much larger than the spacecraft's (approximately rectangle: $I_x$=6000 kg·m², $I_y$= $I_z$ =7687.5 kg·m²).

According to the above mission requirements, a preliminary design of the attitude & orbit control subsystem of the AKI is performed. Control Moment Gyroscope (CMG) and the thrusters are selected as the actuators of the attitude & orbit control subsystem. The CMG realizes the absorption of external disturbance torque through angular momentum exchange with the spacecraft, and keeps the spacecraft attitude stable. At the same time, it can also change the spacecraft attitude quickly to realize the rapid pointing of the payload to the target area. The torque range that a CMG output is $10^{-2}$~$10^3$ N·m.

The thrusters are mainly used for CMG unloading and the attitude & orbit control in the terminal guidance stage. In order to solve the problem that the center of mass is located on the launch vehicle final stage, all thrusters are installed symmetrically on the propulsion module of the spacecraft. At the same time, in order to maximize the arm of force, attitude control is performed through the cooperation of the thrusters installed on the diagonal direction. 3-axis control capability can be achieved by this distribution of the thrusters. Figure 5 shows the view of the propulsion module of the AKI.

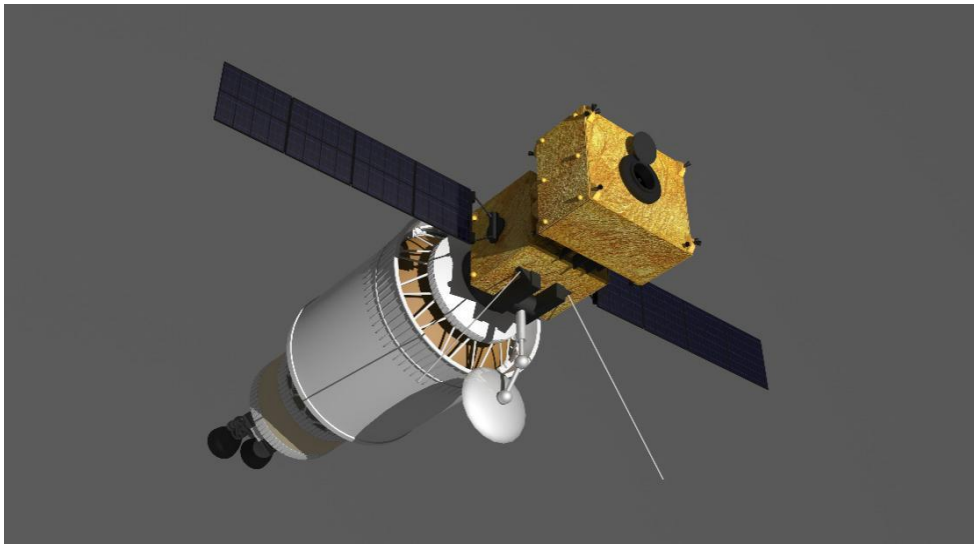

Figure 5 The view of the propulsion module of the Assembled Kinetic Impactor

The details of the thruster distribution are described as follows: monopropellant thrusters are proposed to be used on the propulsion module, with a total of 12 thrusters in 6 groups. Figure 6 shows a schematic diagram of the thruster distribution. No. 1~8 are the thrusters for both orbit control and attitude control. In order to prevent the thruster plumes from affecting the solar arrays and launch vehicle final stage, they are installed symmetrically on the 8 top corners of the propulsion module, the thrusters point at an angle of 30° (30° is according to the engineering experience of space missions) with the roll-axis of the AKI, which are symmetrically distributed along the diagonal. No. 9~12 are the thrusters for attitude control. In order to visualize the angular relationship between the thrusters, Figure 7 shows the distribution of thrusters in different views, where y, z axes are along the diagonal direction and x axis is along the roll-axis of the AKI.

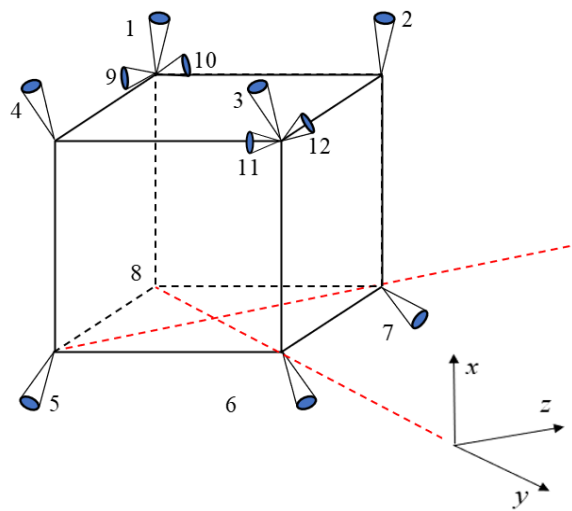

Figure 6 The schematic diagram of the thruster distribution (3-dimensional)

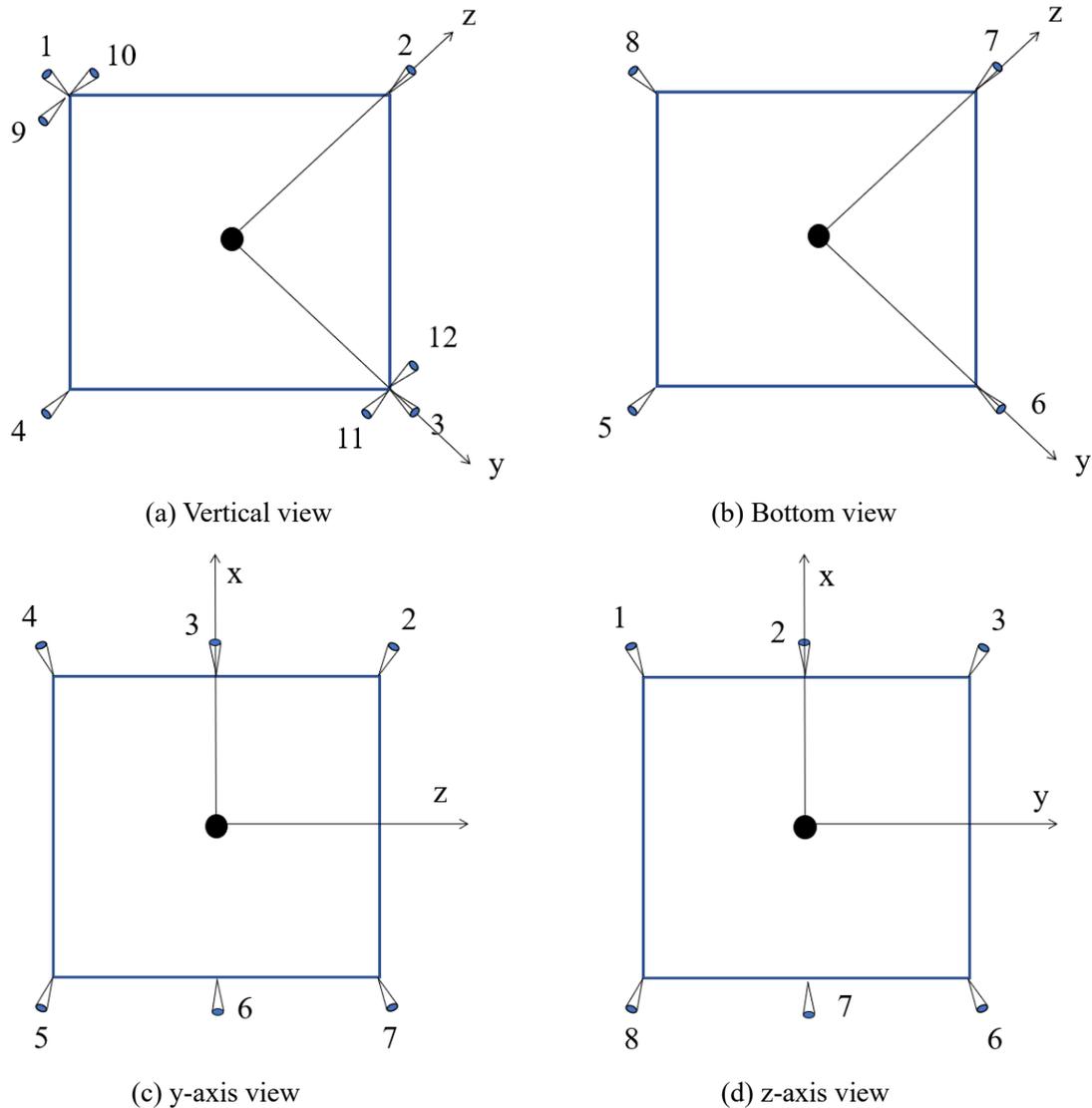

(a) Vertical view  (b) Bottom view

(c) y-axis view  (d) z-axis view

Figure 7 Thruster distribution in different views

Attitude control strategy: pitch channel (y-axis) is controlled by thrusters No. 4&7 or No. 2&5, yaw channel (z-axis) is controlled by thrusters No. 1&6 or No. 3&8, roll channel (x-axis) is controlled by thrusters No. 9&12 or No. 10&11, acceleration channel is controlled by thrusters No. 6&8 or No. 5&7, deceleration channel is controlled by thrusters No. 1&3 or No. 2&4.

The above distribution of the thrusters has the following main advantages: 1) the attitude control and orbit control can be co-used while the coupling of each channel can be eliminated (When the orbit control is performed, the thrust component along the y, z axes can be cancelled each other. When the attitude control is performed, force couples around the x, y, z axes can be formed); 2) the thruster plumes will not affect the solar arrays and launch vehicle final stage. Figure 8 shows the analysis diagram of the thruster plumes.

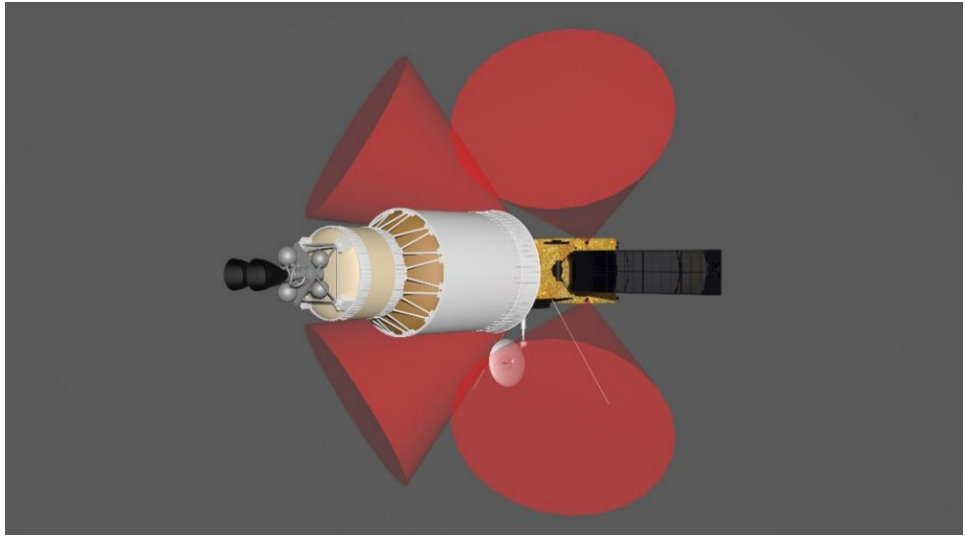
Figure 8 Analysis diagram of the thruster plumes

## Methods

This chapter will mainly introduce the trajectory optimization method of the Assembled Kinetic Impactor (AKI) mission (the optimization method of CKI mission is same to AKI mission). The AKI's transfer model is shown in Figure 9. Assuming the AKI is launched directly into the impact orbit without any Δ*v* during the transfer trajectory, the mass of the AKI at the time of impact is equal to the mass at the time of departure (spacecraft mass and launch vehicle final stage mass).

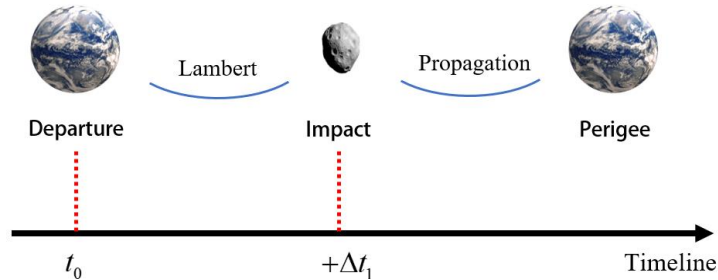

Figure 9 AKI transfer model. The AKI is launched from Earth at $t_0$, impact the asteroid after $\Delta t_1$ days. By solving the Lambert problem, the hyperbolic excessive velocity $\boldsymbol{v}_\infty$ at Earth's SOI and the impact velocity $\boldsymbol{v}_{AKI}$ can be calculated. The $\boldsymbol{v}_\infty$ can be provided by launch vehicle, and the corresponding spacecraft mass of the AKI can be calculated by launch capability.

Assuming that the impact process is a complete inelastic collision with two spheres, according to the law of conservation of momentum, the velocity increment of the asteroid caused by the impact is

$$\Delta \boldsymbol{v}_{Ast} = \beta \frac{m_{AKI}}{m_{AKI} + m_{Ast}} (\boldsymbol{v}_{AKI} - \boldsymbol{v}_{Ast})$$

The deflected asteroid trajectory is propagated by using a Runge-Kutta-Fehlberg 7(8) numerical method to numerically integrate the first-order form of the orbital equations of motion, operating on an N-body solar system dynamics model that includes the gravity of 8 planets and the Moon, the General Relativistic (GR) effect of the Sun, Solar Radiation Pressure (SRP) and Yarkovsky effect. The second-order heliocentric equation of motion is given by

$$\frac{d^2\mathbf{r}}{dt^2} = -\frac{\mu_s}{|\mathbf{r}|^3}\mathbf{r} - \sum_{i=1}^{8}\mu_{pi}\left(\frac{1}{|\mathbf{d}_{pi}|^3}\mathbf{d}_{pi} + \frac{1}{|\boldsymbol{\rho}_{pi}|^3}\boldsymbol{\rho}_{pi}\right) + \mathbf{a}_{moon} + \mathbf{a}_{GR} + \mathbf{a}_{SRP} + \mathbf{a}_{Yarkovsky}$$

Where $\mathbf{d}_{pi}$ indicates the vector from $i^{th}$ planet to the asteroid, $\boldsymbol{\rho}_{pi}$ indicates the vector from $i^{th}$ planet to the Sun. The lunar and planetary ephemerides are based on JPL DE430[16]. The asteroids ephemerides are downloaded from JPL Horizons On-Line Ephemeris System.

According to the above transfer model, there are two decision variables in the optimization model $X = [t_0, \Delta t_1]$. To maximize the asteroid's minimum geocentric distance during the close encounter, the objective function can be described as

$$J = \|\Delta \mathbf{r}'(t_p')\| - \|\Delta \mathbf{r}(t_p)\|$$

where $\Delta \mathbf{r}'(t_p')$ and $\Delta \mathbf{r}(t_p)$ indicates the asteroid's perigee vector after and before the deflection. The genetic algorithm (GA)[17] is an evolutionary computational technique. Because of its global optimization ability, GA is used for the AKI mission optimization in this paper. Figure 10 shows the optimization process of an AKI mission.

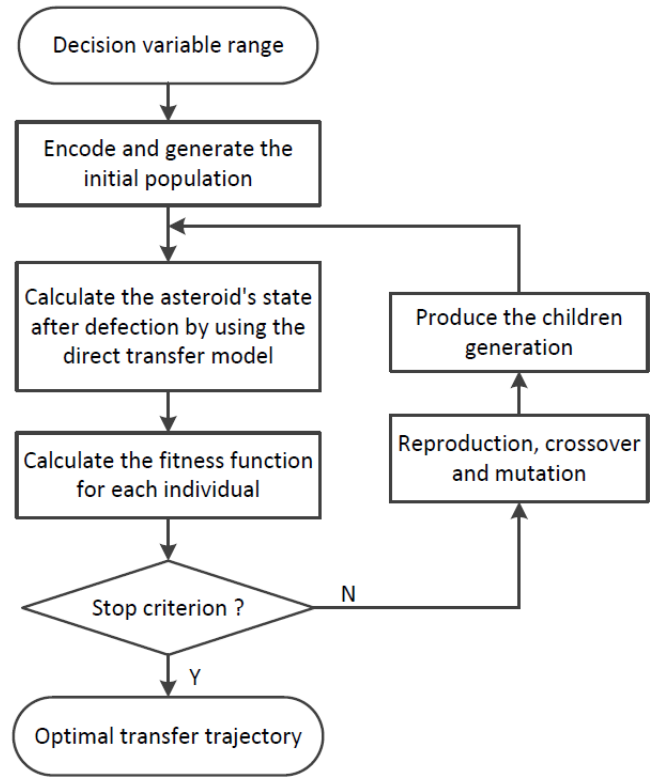

Figure 10 Optimization process

## Conclusions

In order to improve the efficiency of the kinetic impactor strategy, this paper proposes the Assembled Kinetic Impactor (AKI) via combining the spacecraft and the launch vehicle final stage. Based on the technical data of Long March 5 (CZ-5), the missions of deflecting Bennu are designed to demonstrate the power of the AKI. According to the mission design results, the technical feasibility of the AKI is preliminarily studied. The simulation results show that, compared with the Classic Kinetic Impactor (CKI), the AKI concept can greatly improve the deflection efficiency, reduce the launch cost and reduce the launch lead-time. The AKI concept makes it possible to defend Bennu-like large asteroids by a no-nuclear technique within 10 years. At the same time, for a single CZ-5, the deflection distance of a 140 m diameter asteroid within 10 years, can be increased from less than 1 Earth radii to more than 1 Earth radii.

# Acknowledgments

This research was supported by the Beijing Municipal Science and Technology Commission (Z181100002918004)，Civil Aerospace pre research project (D020302，D020304), Strategic Priority Program on Space Science (XDA1502030502, XDA15014900). The corresponding author，who is supported by Youth Innovation Promotion Association CAS, thank the support of CAS Interdisciplinary Innovation Team (JCTD-2018-11). We thank Keqi Li for his contribution in drawing the schematic diagrams, Jianzhao Ding from Beijing Institute of Satellite Control Engineering for the valuable discussion on the orbital attitude control system, Chunping Zeng from DFH satellite Co., Ltd. for the useful discussion on the spacecraft system.

## Author Contributions

Mingtao Li proposed Assembled Kinetic Impactor (AKI) concept and guided the demonstration work of the AKI. Yirui Wang completed the simulations of the AKI mission and finished the paper. Zizheng Gong checked the feasibility of the AKI concept. Jianming Wang provided information about the Long March 5. Chuankui Wang provided the information about the orbit control. Binghong Zhou participated in the demonstration of the AKI.

## Additional Information

**Competing Interests:** The authors declare no competing interests